\documentclass[sn-mathphys-num]{sn-jnl}

\usepackage{graphicx}%
\usepackage{multirow}%
\usepackage{amsmath,amssymb,amsfonts}%
\usepackage{amsthm}%
\usepackage{mathrsfs}%
\usepackage[title]{appendix}%
\usepackage{xcolor}%
\usepackage{textcomp}%
\usepackage{manyfoot}%
\usepackage{booktabs}%
\usepackage{listings}%

\usepackage{lmodern}

\usepackage{bm}
\usepackage{mathrsfs,dsfont}
\usepackage{color}
\usepackage{ulem}
\usepackage{comment}

\newcommand{\bra}[1]{\langle #1 |} 
\newcommand{\ket}[1]{| #1 \rangle } 
\newcommand{\tr}{\mathrm{tr}}
\newcommand{\Schr}{{Schr\"{o}dinger}}
\newcommand{\Ho}{\hat{H}}
\newcommand{\To}{\hat{T}}
\newcommand{\muo}{\hat{\mu}}
\newcommand{\ro}{\hat{\rho}}

\newcommand{\Phimf}{\Phi_\mathrm{mf}}
\newcommand{\VoG}{\hat{V}_G}
\newcommand{\VoGmf}{\VoG^\mathrm{mf}}

\begin{document}
\title[A healthier stochastic semiclassical gravity: world without \Schr\  cats]
          {A healthier stochastic semiclassical gravity: world without \Schr\ cats}
\author[1,2]{\fnm{Lajos} \sur{Di\'osi}}\email{diosi.lajos@wigner.hu}
\affil[1]{\orgname{Wigner Research Center for Physics}, \orgaddress{\street{Konkoly Thege 25-27}, \city{Budapest}, \postcode{H-1225}, \country{Hungary}}}
\affil[2] {\orgname{E\"otv\"os Lor\'and University}, \orgaddress{\street{stny}, \city{Budapest}, \country{Hungary}}}

\abstract{
Semiclassical gravity couples classical gravity to the quantized matter in meanfield approximation.
The meanfield coupling is problematic for two reasons. First, it ignores the quantum fluctuation of matter distribution.
Second,  it violates the linearity of the quantum dynamics. The first problem can be be mitigated by
allowing stochastic fluctuations of the geometry but the second problem lies deep in quantum foundations. 
Restoration of quantum linearity requires a conceptual approach to hybrid classical-quantum coupling.
Studies of  the measurement problem and the quantum-classical transition point the way to a solution.
It is based on a postulated mechanism of spontaneous quantum monitoring plus feedback.
This approach eliminates \Schr\ cat states, takes quantum fluctuations into the account, 
and restores the linearity of quantum dynamics. 
Such conceptionally `healthier' semiclassical theory is captivating,
exists in the Newtonian limit, but its relativistic covariance hits a wall.
Here we will briefly recapitulate the concept and  its realization in the nonrelativistic limit.
We emphasize that the long-known obstacles to the relativistic extension lie in quantum foundations.
}


\maketitle

\section{Introduction}

Quantum theory was invented for the microscopic world, and
proved accurate there. Is it valid in the macroscopic world
as well? Is quantum theory universal from particle physics to
cosmology? We might like to think so. Except that the experimental
evidences are lacking, the relevant quantum gravity theories limp along, 
and there are crippling conceptual problems.  Maybe we cannot quantize
gravity because it does not need to be quantized: space-time is classical.
Then Einstein classical metrics would interact with the quantized
fields of matter. This raises the problem of hybrid classical-quantum
coupling. The prototype of such hybrid dynamics is the standard semiclassical
gravity  \cite{moller1962,rosenfeld1963}, based on the semiclassical Einstein equation:
\begin{equation}\label{EE}
G_{ab}(x)=8\pi G\bra{\Psi}\To_{ab}(x)\ket{\Psi}.
\end{equation}
The curvature of the space-time is sourced by the quantum expectation value ---the meanfield---
of the energy-momentum operator of the matter. 
The fluctuations of $\To_{ab}$ are ignored, do not back-react on the space-time
geometry but the mean values do.  Standard semiclassical gravity is plagued by 
fundamental anomalies because the meanfield coupling violates the obligate 
linearity of quantum dynamics (see in Sec. 2).

The stochastic semiclassical gravity \cite{martin1999,hu2008} takes lowest order quantum 
fluctuations into the account. It mimics them by the zero-mean stochastic field $\delta T_{ab}$,
defined by the following quantum correlator
\begin{equation}\label{TabTcdQFT}
\mathsf{E}\delta T_{ab}(x)\delta T_{cd}(y) = \mathrm{Herm}\bra{\Psi}\To_{ab}(x)\To_{cd}(y)\ket{\Psi}
                                                                                    -\bra{\Psi}\To_{ab}(x)\ket{\Psi}\bra{\Psi}\To_{cd}(y)\ket{\Psi}.
\end{equation} 
Then the modified semiclassical equation
\begin{equation}\label{stosclEE}
G_{ab}=8\pi G\left(\bra{\Psi}\To_{ab}\ket{\Psi}+\delta T_{ab}\right)
\end{equation}
implies  stochastic fluctuations $\delta g_{ab}$ of the metrics as well.
Unfortunately, this stochastic semiclassical gravity  does not mitigate 
the fundamental anomalies of the semiclassical Einstein equation.

Reviving the concept of spontaneous quantum monitoring plus feedback  \cite{diosi1990},  
we proposed the `conceptionally healthier semiclassical' gravity \cite{tilloydiosi2016sourcing,tilloydiosi2017principle}
which is free of the said anomalies. It ensures the obligate linearity of quantum dynamics  
since it is based on standard measurement-plus-feedback mechanisms. Accordingly,
one assumes  that the energy-momentum operator $\To_{ab}$ is universally monitored,
i.e.: measured everywhere and every time.
The monitored (measured) value is a classical random tensor field:
\begin{equation}\label{Tabsignal}
T_{ab}=\bra{\Psi}\To_{ab}\ket{\Psi}+\delta T_{ab}
\end{equation}
with stochastic fluctuation  $\delta T_{ab}$ around the meanfield. This monitored value
is then used for feedback on the r.h.s. of the classical Einstein equation. 
This $T_{ab}$  coincides formally with the expression in
the phenomenological equation (\ref{stosclEE}) of stochastic semiclassical gravity but the difference is crucial. 
The stochastic term is  defined differently.
It is just the measurement noise and its correlation is determined by the chosen 
precision and correlations of the local measurements constituting the monitoring setup: 
\begin{equation}\label{TabTcd}
\mathsf{E}\delta T_{ab}(x)\delta T_{cd}(y) = D_{ab|cd}(x,y).
\end{equation}
The r.h.s. must be a covariant translation invariant non-negative kernel.
Our proposal differs from the phenomenological stochastic theory \cite{martin1999,hu2008} in a second
way, too. The dynamics of the state $\Psi$ must contain the non-Hamiltonian influence of the monitoring.
However promising this project was, refs. \cite{diosi1990,tilloydiosi2016sourcing,tilloydiosi2017principle} 
could have only realized the Newtonian non-relativistic limit,. with the clear identification of the 
obstacle to the relativistic version. 
Interestingly, the formalism of the `healthier' semiclassical dynamics was recently applied
relativistically, see \cite{oppenheim2023postquantum,layton2024healthier} and refs. therein. 
The so-called postquantum gravity \cite{oppenheim2023postquantum} would be
a `healed' semiclassical gravity, consistent with quantum theory.   

This work proposes a tour. We start from the standard semiclassical
gravity and then we `descend' to its non-relativistic Newtonian limit. 
This is the best way to identify the fundamental quantum  anomalies of the
meanfield coupling and their resolution by the `healthy' 
stochastic modification postulating the mechanism of spontaneous quantum
monitoring and feedback.  
Then we try to `ascend' to the relativistic realization
but we will find the old obstacle that we  had known before.

\hskip18pt
\begin{center}
\includegraphics[width=.85\textwidth]{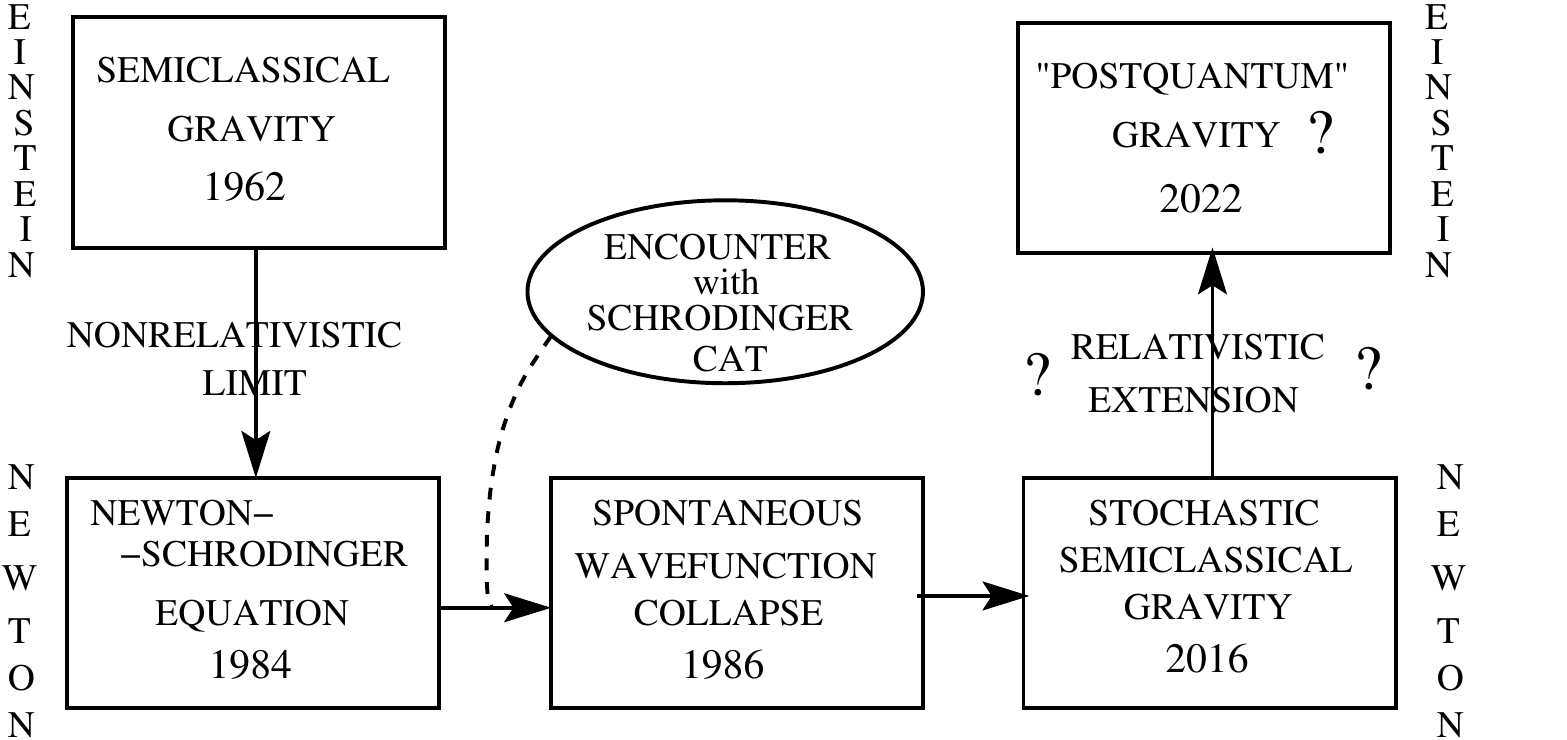}
\end{center}

\section{Semiclassical Gravity}
Consider a given foliation of the space-time in spacelike hypersurfaces $\Sigma$ and the 
\Schr~ state vector $\ket{\Psi_\Sigma}$ on it. 
The classical metric $g_{ab}$ which will be the solution of
the semiclassical Einstein equation \eqref{sclEE}.
The state vector of the quantized matter evolves with the Tomonaga--Schwinger equation \eqref{TSE}
where the Hamiltonian density $\hat{\mathcal{H}}$ depends on the solution  $g_{ab}$
which is the $\Psi$-dependent solution of eq. \eqref{sclEE} causing the non-linearity of eq. \eqref{TSE}:
\begin{eqnarray}
\label{sclEE}
G_{ab}(x)&=&8\pi G\bra{\Psi_\Sigma}\hat{T}_{ab}(x)\ket{\Psi_\Sigma},~~~~~~(x\in\Sigma),\\
\label{TSE}
\frac{\delta\ket{\Psi_\Sigma}}{\delta\Sigma(x)}&=&-i\hat{\mathcal{H}}(x)\ket{\Psi_\Sigma}.
\end{eqnarray}
The semiclassical gravity  \cite{moller1962,rosenfeld1963} is a powerful hybrid dynamics
of classical gravity and quantized matter.
In the Newtonian limit the semiclassical eqs. (\ref{sclEE},\ref{TSE}) become much simpler: 
\begin{eqnarray}
\label{Phimf}
\Phi(r,t)&=&\Phimf(r,t)=-G\int\bra{\Psi_t}\muo(r')\ket{\Psi_t}\frac{d^3r'}{\vert r-r'\vert},\\
\label{SNE}
\frac{d\ket{\Psi_t}}{dt}&=&-i\left(\Ho+\int\muo(r)\Phimf(r,t)d^3r\right)\ket{\Psi_t}=\nonumber\\
                                            &=&-i\left(\Ho-G\int\int\muo(r)\bra{\Psi_t}\muo(r')\ket{\Psi_t}\frac{d^3r d^3r'}{|r-r'|}\right)\ket{\Psi_t}
\end{eqnarray}
where 
$\Phimf$ is the mean-field Newton potential, $\mu$ is the distribution of non-relativistic mass density and
$\Ho$ is the self-Hamiltonian. Observe that the semiclassical equation (\ref{SNE}) of non-relativistic
quantized matter, called the \Schr--Newton equation, does not contain the trivial Newton pair-potential
\begin{equation}\label{VoG}
\VoG=-G\int\int\muo(r)\muo(r')\frac{d^3r d^3r'}{\vert r-r'\vert}.
\end{equation}
Instead,  it contains a non-linear meanfield term $\VoGmf$ to represent gravity's back-reaction:
\begin{equation}\label{VoGmf}
\VoGmf=-G\int\int\muo(r)\langle\muo(r')\rangle\frac{d^3r d^3r'}{\vert r-r'\vert}.
\end{equation}
Note incidentally, that
the eq. (\ref{SNE}) had already been used  for quantized stellar masses \cite{ruffini1969systems}.
For long time,  impressed by relativistic field theories, its relevance at low energies has remained overlooked.     
It was revealed finally \cite{diosi1984,penrose1996} that both gravity and quantumness might become 
relevant together non-relativistically for massive degrees of freedoms,  e.g., in center-of-mass motion of 
of objects with masses of the order of nanograms. 

The Newtonian limit (\ref{Phimf},\ref{SNE}) not only perpetuates  the fundamental 
anomalies of semiclassical gravity (\ref{sclEE},\ref{TSE})
but also understands the causes \cite{diosi2016nonlinear}. Core problems are 
fake-action-at-a-distance (aka causality violation relativistically) and the breakdown
of Born's statistical interpretation of the wavefunction $\Psi$.
They are caused by the $\Psi$-dependent meanfield  potential $\Phimf$ in the \Schr--Newton equation (\ref{SNE}).
The meanfield coupling $\bra{\Psi}\muo\ket{\Psi}$ should be blamed.

What else should we use? It is necessary that we consider the possible hybrid classical-quantum (CQ)
couplings in the light of quantum foundations. Action of C on Q is parametric and makes
no problem. Back-reaction of Q on C is the major issue. Still, the answer is there in the fundaments
of quantum theory.  About an \textit{individual} quantum system, quantum measurement is the only 
way to consistently define classical variables. Classical numbers like $\bra{\Psi}\muo\ket{\Psi}$
are not classical variables, their  coupling  to classical systems is illegitimate.  
Only the random measurement outcomes  are legitimate classical variables suitable to couple to
other classical variables.
A lesson is important here. Reversibility of hybrid systems is lost
for two reasons. First, measurement imposes decoherence on Q. Second, coupling to
the random measurement outcome imposes stochasticity of C. 

So, back-reaction of quantized matter on classical gravity is only possible via the 
random measurement outcomes $\mu$ of $\muo$-monitoring, instead of 
the meanfield $\bra{\Psi}\muo\ket{\Psi}$. The measured outcomes $\mu(r,t)$
contain the mean values plus the measurement noise:
\begin{equation}\label{mu}
\mu(r,t)=\bra{\Psi_t}\muo(r)\ket{\Psi_t}+\delta\mu(r,t),
\end{equation}
where $\delta\mu$ is a white noise with possible spatial correlations. 
Accordingly, the calculated Newton potential, too, contains an additional white noise:
\begin{equation}\label{dPhi}
\delta\Phi(r,t)=-G\int\delta\mu(r',t)\frac{d^3r'}{\vert r-r'\vert}. 
\end{equation}
In the \Schr--Newton eq. (\ref{SNE}), we shall  replace the meanfield $\Phimf$ by $\Phimf+\delta\Phi$
for reconciliation of the stochastically modified
\Schr--Newton semiclassical equation with standard quantum theory. But first we need to answer
two questions. Who is measuring (monitoring) the mass density $\muo$? That's the truly sensitive
question. The other one is more technical: how to parametrize the $\muo$-monitoring, i.e., 
how to choose the spatial correlations of $\delta\mu$. 
Towards answering both questions, sec. 3  recalls the postulate of single spontaneous collapse and 
its gravity-related parametrization. This single spontaneous collapse
 is then upgraded into the postulate of spontaneous monitoring of the mass density $\mu$.
 Monitoring means time-continuous measurement.
 Spontaneity means that no instruments (no observers) are present while the dynamics is
 undergoing the same standard stochastic modifications as if they were there. 
 
\section{Spontaneous collapse of \Schr\ Cats}
If we extend the validity of quantum theory for large masses, and this is what we do if we  believe in
quantum cosmology, then we are faced with some counterintuitive situations. The paradigmatic
one is the \Schr~ cat state. Consider the quantized center-of-mass motion of a macroscopic mass $M$ prepared
in a balanced superposition on the `left' and `rigth' respectively, with macroscopic distance from each other. 
The existence of such a state is trivial in the nontrivial manyworld
interpretation of the quantum theory, at least paradoxical in conservative quantum theory, 
and viewed nonsense by some. Without taking side, one can speculate about a quantum theory 
that is free of such massive macroscopic superpositions. 
We postulate the collapse of the macroscopic superposition
as if it would happen under a measurement, to happen spontaneously this time:
\begin{equation}
\ket{CAT}=\frac{\ket{\mbox{LEFT}}+\ket{\mbox{RIGHT}}}{\sqrt{2}}
\rightarrow\left\{\begin{array}{c}\ket{\mbox{LEFT}}\\
                                                                 \mbox{or}\\
                                                                 \ket{\mbox{RIGHT}}
                                  \end{array}\right. .
\end{equation}
We should propose a collapse time. 
Let us follow the proposal of Penrose and the present author.
At this point gravitation and \Schr~ cats encounter.
Quantized massive objects in spacetime lead to controversies with sharply defined Newtonian potential
non-relativistically \cite{diosi1987} and with sharply defined time-flow relativistically \cite{penrose1996}. 
Therefore we allow a certain unsharpness
$\delta\Phi$ of the Newton potential (of the time-flow, relativistically). If this unsharpness is represented
stochastically by a white noise then, 
despite their different explanations, the proposals in refs. \cite{diosi1987} and \cite{penrose1996}, resp., 
correspond to the  same measure of unsharpness: 
\begin{equation}\label{dPhidPhi}
\mathsf{E}\delta\Phi(r,t)\delta\Phi(r',t')=\frac{\hbar G/2}{\vert r-r'\vert}\delta(t-t'),
\end{equation}
Both authors derive that this unsharpness leads to the following characteristic rate of the collapse:
\begin{equation}\label{DPtau}
\frac{1}{\tau}=\frac{V_G^i-V_G^f}{\hbar},
\end{equation}
where $V_G^i$ and $V_G^f$ are formally identified as the classical Newtonian pair-potential between two copies of the
mass M in the separate and the coincident positions, respectively.
Now we see that the postulated spontaneous collapse is ignorable in the microscopic degrees of freedom but
it becomes dominant gradually for large masses. The collapse rate, still negligible ($\sim10^{-6}/s$) for a femtogram, 
is overwhelming fast ($\sim10^6/s$) for a milligram mass already. 

The non-relativistic theory \cite{diosi1987,diosi1989} and \cite{penrose1996}, 
based on the above postulate of gravity-related spontaneous collapse rates, 
is a possible explanation of the classical-quantum transition, a world without \Schr~ cats and a theory
without the measurement problem.  The next section upgrades it into the consistent non-relativistic hybrid dynamics
of quantized matter and classical gravity.

\section{On the healthier semiclassical gravity}
As Sec. 2 anticipated, the meanfield Newton potential (\ref{Phimf}) is completed by  a noise term:   
\begin{eqnarray}\label{stoPhi}
\Phi(r,t)&=&-G\int\bigl(\bra{\Psi_t}\muo(r')\ket{\Psi_t}+\delta\mu(r',t)\bigr)\frac{d^3r'}{\vert r-r'\vert}=\nonumber\\ 
               &=&\Phimf(r,t)+\delta\Phi(r,t).
\end{eqnarray} 
This corresponds to the Newton field sourced by the spontaneously monitored value 
$\bra{\Psi_t}\muo(r)\ket{\Psi_t}+\delta\mu(r,t)$ of the mass density. 
Sec. 3 proposed the expression (\ref{dPhidPhi}) for the statistics of the noise and we repeat it here for completness:
\begin{equation}\label{dPhidPhi1}
\mathsf{E}\delta\Phi(r,t)\delta\Phi(r',t')=\frac{\hbar G/2}{\vert r-r'\vert}\delta(t-t').
\end{equation}
If in the \Schr--Newton eq. (\ref{SNE}) we replace the meanfield $\Phimf$ by the expression (\ref{stoPhi})
then we obtain the following:
\begin{eqnarray}\label{stoSNE}
\frac{d\ket{\Psi_t}}{dt}&=&-\frac{i}{\hbar}\left(\Ho_0+\VoGmf
+\int\muo(r)\delta\Phi(r,t))d^3r\right)\ket{\Psi_t}
\nonumber\\
&&-\frac{G}{2\hbar}\int\int \frac {\left(\muo(r)-\langle\muo(r)\rangle_t\right)
                                                                \left(\muo(r')-\langle\muo(r')\rangle_t\right)}
                                                                {\vert r-r'\vert}d^3r d^3r'\ket{\Psi}\nonumber\\
&&+\frac{1}{\hbar}\int\left(\muo(r)-\langle\muo(r)\rangle_t\right)\delta\Phi(r,t)d^3r\ket{\Psi_t}.                                     
\end{eqnarray}
The \Schr--Newton equation (\ref{SNE}) has become modified in three ways.
The first line results from the back-reaction engineered by $\Phimf+\delta\Phi$. 
The second line contains the decoherence term because the $\muo$-monitoring causes dynamic suppression of $\muo$'s quantum 
fluctuations. The third line contains a further term with $\delta\Phi$ and represents the random effect of monitoring.
We emphasize that whence the correlation (\ref{dPhidPhi1}) is fixed, 
the postulated spontaneous monitoring leads uniquely to  this result (\ref{stoSNE}) 
via the standard calculus of quantum monitoring and feedback 
\cite{tilloydiosi2016sourcing,tilloydiosi2017principle}. 

Although the `healthier' semiclassical equations (\ref{stoPhi}-\ref{stoSNE}) are consistent by construction 
with quantum theory,
there is an equivalent formalism where the linearity is restored explicitly. 
If by  the relatioship $\Phi(r,t)=d\chi_t(r)/dt$ we introduce the field $\chi$, we can see that
$\chi_t$ and $\Psi_t$ undergo correlated diffusions respectively in the functional and Hilbert space.
Namely, $d\chi/dt$ is driven by the white noise $\delta\Phi$ according to eq. \eqref{stoPhi}, meaning its
diffusion in the functional space, and $d\ket{\Psi_t}/dt$ is also driven by $\delta\Phi$ according to eq. \eqref{stoSNE}, 
meaning its diffusion in the Hilbert space. 
Therefore the couple $(\chi,\Psi)$ can be described by the hybrid of the Fokker--Planck and Lindblad equations.
Using the hybrid $\ro[\chi]$ of the probability density $\rho[\chi]=\tr\ro[\chi]$ and
of the density operator $\ro=\int\ro[\chi]d[\chi]$, ref. \cite{diosi2024classical} derives the following 
hybrid master equation from the stochastic equations (\ref{stoPhi}-\ref{stoSNE}): 
\begin{eqnarray}\label{HME_DP}
&&\frac{d\ro[\chi]}{dt}=-\frac{i}{\hbar}[\Ho_0+\VoG,\ro[\chi]]+\\
&&+G\!\int\!\!\!\int\!
\left(-\frac{1}{2\hbar}[\muo(r),[\muo(s),\ro[\chi]]]
          +\mathrm{Herm}(1\!+\!i)\muo(r)\frac{\delta\ro[\chi]}{\delta\chi(s)}
          +\frac{\hbar}{4}\frac{\delta^2\ro[\chi]}{\delta\chi(r)\delta\chi(s)}\right)
\frac{d^3r d^3s}{|r-s|}.\nonumber
\end{eqnarray}
The non-unitary mechanisms are represented by the second line. The first term corresponds to the
suppression of  $\muo$'s `macroscopic' quantum fluctuations, the middle term is responsible
for the back-reaction of $\muo$ on the classical Newton potential, and the third term stands for the
diffusion of the classical Newton potential.
The  most important new feature of the above hybrid master equation
is the appearance of the Newton pair-potential (\ref{VoG}) in place of the non-linear 
meanfield term $\VoGmf$ (\ref{VoGmf}) in the \Schr--Newton eq. (\ref{SNE}).

The formal relativistic counterpart of either the stochastic eqs.  (\ref{stoPhi}-\ref{stoSNE}) or the equivalent master equation
(\ref{HME_DP}) can obviously be constructed. The stochastic ones would look like this:
\begin{eqnarray}
\label{stoEEXXX}
G_{ab}(x)&=&8\pi G\left(\bra{\Psi_\Sigma}\hat{T}_{ab}(x)\ket{\Psi_\Sigma}+\delta T_{ab}(x)\right),~~~~~~(x\in\Sigma),\\
\label{Dabcd}
\mathsf{E}\delta\To_{ab}(x)\To_{cd}(y)&=&D_{ab\vert cd}\delta(x-y),\\
\label{TSEXXX}
\frac{\delta\ket{\Psi_\Sigma}}{\delta\Sigma(x)}&=&-i\hat{\mathcal{H}}(x)\ket{\Psi_\Sigma}
+\left\{\begin{array}{c}\mbox{nonlinear}\\
                                                     \mbox{stochastic}\end{array}\right\}\mbox{terms of monitoring}.
\end{eqnarray}
Such might be the structure of full relativistic equations of the `healed' stochastic semiclassical gravity,
The equivalent hybrid master equation of classical metrics and the quantized relativistic matter
would contain the three terms: decoherence term to suppress $\To_{ab}$'s large quantum
fluctuations, back-reaction of $\To_{ab}$ on the metrics $g_{ab}$, and the term for the diffusion of $g_{ab}$.
The formal construction of the related `postquantum' gravity faces multiple issues
(e.g.: ensuring diffeomorfism invariance,  renormalizability of divergences)
recognized immediately \cite{layton2024healthier,oppenheim2023postquantum}. As we mentioned,
the theory follows uniquely if the correlation (\ref{Dabcd}) has been chosen. The issues
culminate right here. The covariant correlation must contain the four-dimensional 
$\delta(x-y)$ and it leads to divergences. No covariant regularization and renormalization methods
are available.

The obstacles are not simply technical ones, they are multiply rooted in foundations.
Wavefuntion collapse depends on the reference frame hence 
the covariant formalism of selective quantum measurements and quantum monitoring is problematic.
On the classical end,  diffusion cannot be made
relativistic,  nor even special relativistic \cite{diosi2024classical}. 

\section{Closing remarks}
Unified theory of space-time with quantized matter and the physics
of quantum measurement were considered unrelated for long time,
studied by two separate research communities. Quantum cosmologists
used heavy artillery of mathematics. 
Quantum measurement problem `solvers'-, with the present author
among them, used light weapons and sometimes whimsical identification
of their problems, e.g. in terms of the \Schr\  cat paradox. The               
bottle-neck of quantum gravity may be this paradox, not the math difficulties 
to find a good framework of quantization. An improved
but still semiclassical theory might be based on the non-relativistic
theory of spontaneous quantum monitoring and feedback, eliminating \Schr\ 
cat states. Such healthier theory exists non-relativistically
but its relativistic  - even Lorentzian - extension remains a problem.

It seems that the consistent hybrid theory  of quantized matter and classical gravity
based on relativistic calculus of monitoring plus feedback was first discussed
longtime ago \cite{diosi1990},  with the warning that the relativistic calculus may pose
serious problems. Much later, the authors of refs. \cite{tilloydiosi2016sourcing,tilloydiosi2017principle} 
have repeatedly argued that the `healthier' relativistic semiclassical gravity is blocked
by the continued lack of a relativistic model for quantum monitoring.
The recent proposal of `postquantum' gravity  
(\cite{oppenheim2023postquantum,layton2024healthier} and refs. therein)
is optimistic about the future resolution of its difficulties but it is not conscious
of where the difficulties stem from. 
Refs. \cite{diosi2024classical,tilloy2024general}
point out in strict accordance with  the previous warnings 
\cite{diosi1990,tilloydiosi2016sourcing,tilloydiosi2017principle} that 
the ultimate difficulties of the `healthier' relativistic hybrid dynamics are 
difficulties of relativistic quantum monitoring.
It seems that as long as this latter is missing, the `postquantum' gravity cannot be complete.  
Until we discover the theory of relativistic quantum monitoring, if it exists,   
the foundational application of the `healthier' semiclassical dynamics in its present form
remains an unfulfilled promise though by no means finally discarded.

\backmatter

\bmhead{Acknowledgements}
This work is based on my talk at the Lema\^itre Conference 2024, I am indebted to
the conference organizers. My research was supported  by the
National Research, Development and Innovation Office
''Frontline'' Research Excellence Program (Grant No.
KKP133827), by the John Templeton Foundation (Grant 62099),
and by the EU COST Actions (Grants CA23115, CA23130). 


\end{document}